\documentclass[,final]{aipproc}
\layoutstyle{6x9}

\newcommand{\be}{\begin{equation}}
\newcommand{\ee}{\end{equation}}
\newcommand{\ba}{\begin{eqnarray}}
\newcommand{\ea}{\end{eqnarray}}

\begin{document}

\title{Relic gravitational waves and cosmic accelerated expansion}


\author{Germán Izquierdo}{address={Departamento de Física, Universidad Autónoma de Barcelona, \\
08193 Bellaterra (Barcelona) Spain} }

\begin{abstract}
The possibility of reconstructing the whole history of the scale
factor of the Universe from the power spectrum of relic
gravitational waves (RGWs) makes the study of these waves quite
interesting. First, we explore the impact of a hypothetical era
-right after reheating- dominated by mini black holes and
radiation that may lower the spectrum several orders of magnitude.
Next, we calculate the power spectrum of the RGWs taking into
account the present stage of accelerated expansion and an
hypothetical second dust era. Finally, we study the generalized
second law of gravitational thermodynamics applied to the present
era of accelerated expansion of the Universe.

\end{abstract}

\maketitle

\section{Introduction}
The future detection of relic gravitational waves is expected to
provide us with invaluable information about the instant of their
decoupling from other fields, i.e., about $10^{-43}$ seconds after
the Big Bang. The relic gravitational waves (RGWs) are generated
by parametric amplification of the quantum vacuum during the
expansion of the Universe.

The equation governing the evolution of the RGWs is the so-called
Lifshitz equation  $\mu ^{\prime \prime }(\eta )+\left(
k^{2}-\frac{a^{\prime \prime }(\eta )}{ a(\eta )}\right) \mu (\eta
)=0,$ which can be interpreted as the equation of an harmonic
oscillator parametrically excited by the term $a^{\prime \prime
}/a\,$ \cite{lif}. When $k^{2}\gg a^{\prime \prime }/a$, Liftshitz
equation becomes the equation of the simple pendulum and
consequently the amplitude of the wave decreases adiabatically as
$a^{-1}$ in an expanding universe. In the opposite regime,
$k^{2}\ll a^{\prime \prime }/a$, the dominant solution is
proportional to the scale factor and the amplitude remains
constant with the expansion of the universe. This phenomenon is
called \textquotedblleft superadiabatic\textquotedblright\ or
\textquotedblleft parametric\textquotedblright\ amplification of
gravitational waves \cite{grish74}. Another approach to the RGWs
amplification relies on the method of Bogoliubov coefficients and
uses the adiabatic vacuum approximation (for details see, e.g.,
\cite{Allen}). Here we evaluate the number of RGWs created in the
adiabatic vacuum approximation in a universe which experiences
three successive stages of evolution: an initial de Sitter stage
followed by a stage dominated by the radiation and, finally, a
stage dominated by the non-relativistic matter up to the present
day. The power spectrum, defined from the energy density of the
RGWs as $d\rho_g(\omega)=P(\omega) d\omega$, reads \cite{Allen,
grish93} {\small
\begin{equation}
P\left( \omega \right) \sim \left\{
\begin{array}{ll}
0 & \left( \omega (\eta _{0})>2\pi (a_{1}/a_{0})H_{1}\right) , \\
\omega ^{-1}(\eta _{0}) & \left( 2\pi (a_{2}/a_{0})H_{2}<\omega
(\eta _{0})<2\pi (a_{1}/a_{0})H_{1}\right) , \\
\omega ^{-3}(\eta _{0}) & \left( 2\pi H_{0}<\omega (\eta
_{0})<2\pi (a_{2}/a_{0})H_{2}\right) .%
\end{array}%
\right.  \label{espectr}
\end{equation}}

\section{Relic gravitational waves and mini black holes}
As is well known, mini black holes (MBHs) can be created by
quantum tunnelling from the hot radiation and coexist with it
until their evaporation \cite{gross}. It is reasonable to expect
that, at this point, a steady state in the very early Universe
would be achieved where the total energy density is shared between
the black holes and radiation whence $\rho =\rho _{BH}+\rho _{R}$
and, consequently, the total pressure is $ p=p_{R}=(\gamma
-1)\rho,$ where the constant $\gamma$ lies in the interval $1\leq
\gamma <4/3$. If the density of MBHs is large enough to dominate
the expansion of the Universe, then $\gamma \simeq 1 $. In the
opposite case, the Universe expansion is dominated by the
radiation, $\gamma \simeq 4/3$. During the \textquotedblleft
MBHs+rad\textquotedblright\ era ,  one finds from the Einstein
equations that $a(\eta)\propto \eta^l$, where $l=2/(3\gamma-2)$
and $1< l \leq 2$ \cite{MBHs}.

According to this, it seems reasonable to assume a four-stage
model of universe, initially de Sitter, then dominated by a
mixture of MBHs and radiation, then dominated by the radiation
after the evaporation of the MBHs and finally dominated by dust
until today. The only free parameters considered here are $l$, the
duration of the \textquotedblleft MBHs+rad\textquotedblright\ era,
$\tau $, and the Hubble factor in the de Sitter era, $H_{1}$.

\begin{figure}[tbp]
\includegraphics*[height=11cm]{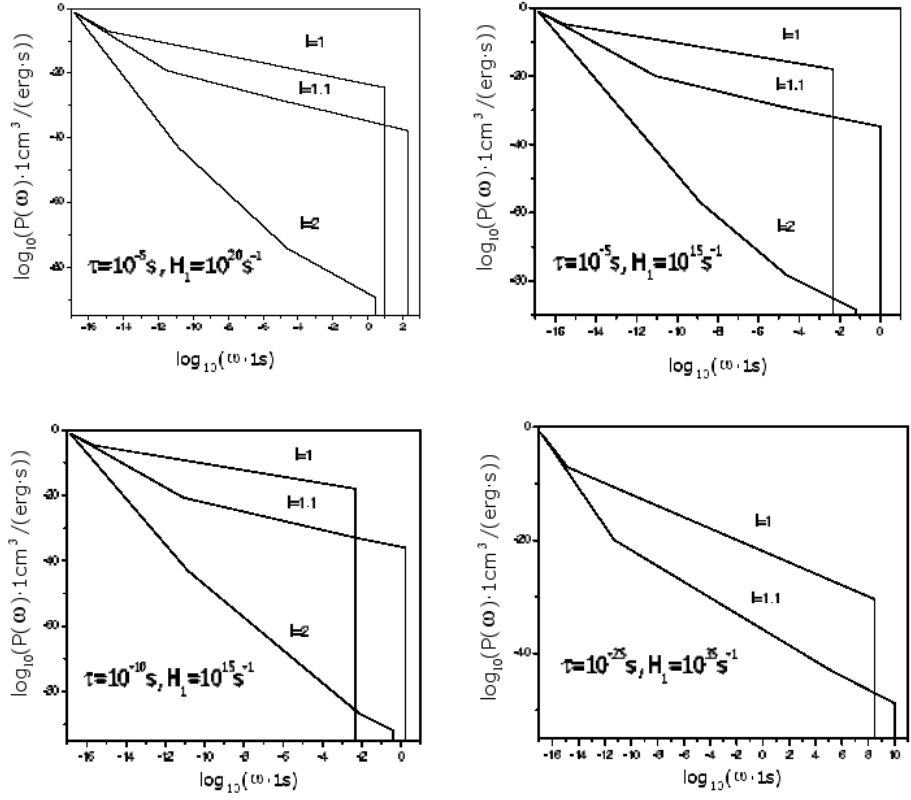}

\caption{GWs spectrum for an expanding universe with a
``MBHs+rad'' era for certain values of $l$, $\protect\tau $ and
$H_{1}$.} \label{fig1}
\end{figure}

The power spectrum for this four-stage scenario is plotted in
figure \ref{fig1}. The spectrum predicted for the three-stage
model of the previous section is shown for comparison (labelled as
$l=1$). Parameters $\tau$ and $H_1$ are chosen assuming that each
spectrum has the maximum value allowed by the CMB anisotropy data
at the frequency $\omega=2\pi H_0=2.24\times 10^{-18}s^{-1}$. The
predicted power spectrum of the four-stage model is lower than the
three-stage one by several orders of magnitude. In the plot of the
bottom-right panel the power spectrum for $l=2$ is excluded as it
predicts a RGWs energy density at the matter-radiation decoupling
that generates via the Sach-Wolfe effect a CMB anisotropy larger
than observed.

\section{Present accelerated expansion}
\begin{figure}
\includegraphics*[height=7cm]{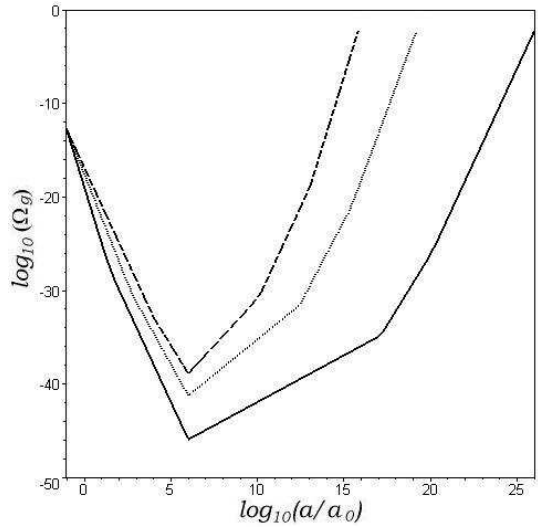}
\caption{ Evolution of the density parameter $\Omega_g$ with the
scale factor in the five-stage scenario (De Sitter
inflation-radiation-dust-dark energy-second dust era) from the
beginning of the dark energy era. The solid, dotted and dashed
lines correspond to $l=-1$, $l=-1.5$ and $l=-2$, respectively.}
\label{fig2}
\end{figure}

Some models of dark energy predict that the present accelerated
phase of cosmic expansion governed by $a(\eta)\propto\eta^l$ with
$l\leq -1$ is transitory and that the expansion, sooner or later,
will be dominated by ``dust''\footnote{These models predict that
the dark energy will evolve in a way that mimics the expansion of
a universe dominated by non relativistic matter.} again
\cite{Alam}. This five-stage model of universe (de
Sitter-radiation era-dust era-dark energy era-``second dust'' era)
predicts a current RGWs power spectrum that is not at variance
with the one of the three-stage scenario but evolves differently.
As the universe expands in the three-stage model the Hubble volume
is continuously increasing and new RGWs are reentering it and
contributing to de power spectrum. Meanwhile, in the dark energy
scenario during the accelerated era the Hubble volume decreases,
the RGWs begin to leave it and cease to contribute to the power
spectrum \cite{RGWPAE}. Once the universe reaches the second dust
era the Hubble radius begins to grow again and the RGWs reenter it
in a $l$ dependent way. Figure \ref{fig2} shows the evolution of
the RGWs density parameter, defined as $\Omega_g=\rho_g/\rho_c$.

\section{Relic Gravitational waves entropy and GSL}

During the present accelerated era of expansion the number density
of RGWs is decreasing with time and, eventually, all the RGWs will
leave the Hubble radius.

It seems reasonable to assume that the entropy of the RGWs depends
on their number present in the considered volume \cite{gasperini}.
Thus, the entropy of the RGWs decreases with time. In the simplest
assumption the entropy density of RGWs is proportional to their
number density, i.e. $s_g=A n_g$. The generalized second law of
gravitational thermodynamics\footnote{According to he generalized
second law (GSL) of gravitational thermodynamics, the entropy of
the event horizon plus its surroundings (in our case, the entropy
in the volume enclosed by the horizon) cannot decrease.} will be
satisfied if the constant $A$ is lower than certain bound, which
is model dependent \cite{RGWsGSL}. This is the only information at
our disposal on this proportionality constant.

\section{Conclusions}

The MBH four-stage scenario predicts a much lower power spectrum
than the conventional three-stage scenario for the same $H_1$. The
free parameters of this scenario are constrained by the CMB
anisotropy data. Although the current power spectrum of the RGWs
in the dark energy four-accelerated scenario is not at variance
with that of the three-stage scenario, the RGWs density parameter
evolves differently. Its future evolution may also help discern
between different dark energy decaying models. Assuming that the
entropy density of the RGWs is proportional to their number
density, the GSL is fulfilled provided a condition over the
proportionality constant is met.


\end{document}